\documentclass[]{aa}  

\usepackage{graphicx}
\usepackage{txfonts}

\usepackage{natbib}
\bibpunct{(}{)}{;}{a}{}{,} 

\usepackage[colorlinks=true,linkcolor=blue, citecolor=blue, filecolor=blue, urlcolor=blue]{hyperref}
\usepackage{orcidlink}

\begin{document}

   \title{Modelling spacecraft-emitted electrons measured by SWA-EAS experiment on board Solar Orbiter mission}
   \titlerunning{Modelling electron emissions from Solar Orbiter spacecraft}

   \author{Š.~ Štverák\inst{1,2}\orcidlink{0000-0002-5870-2043}
          \and          
          D.~Herčík\inst{1}\orcidlink{0000-0001-6151-7482}
          \and
          P.~Hellinger\inst{1,2}\orcidlink{0000-0002-5608-0834}
          \and
          M.~Pop\v{d}\!akunik\inst{1,3}\orcidlink{0009-0007-5632-7458}
          \and
          G.~R.~Lewis\inst{4}\orcidlink{0000-0002-5982-4667}
          \and
          G.~Nicolaou\inst{4}\orcidlink{0000-0003-3623-4928}
          \and
          C.~J.~Owen\inst{4}\orcidlink{0000-0002-5982-4667}
          \and
          Yu.~V.~Khotyaintsev\inst{5}\orcidlink{0000-0001-5550-3113}          
          \and
          M.~Maksimovic\inst{6}\orcidlink{0000-0001-6172-5062}
          }

   \institute{
             Institute of Atmospheric Physics of the Czech Academy of Sciences, Boční II 1a/1401, 14100 Prague, Czechia\\
             \email{stverak@ufa.cas.cz}
             \and
             Astronomical Institute of the Czech Academy of Sciences, Boční II 1a/1401, 14100 Prague, Czechia
             \and
             Czech Technical University in Prague, Prague, Czechia
             \and
             Department of Space and Climate Physics, Mullard Space Science Laboratory, University College London, Dorking, Surrey~RH5~6NT, UK             
             \and
             Swedish Institute of Space Physics (IRF), Uppsala 75121, Sweden
             \and             
             LIRA, Observatoire de Paris, Université PSL, CNRS, Sorbonne Université, Université de Paris, 5~Place Jules Janssen, 92195~ Meudon, France
             }

   \date{Received 7 January 2024 / Accepted 20 March 2026}

   \abstract
   {}
   {Thermal electron measurements in space plasmas typically suffer at low energies from spacecraft emissions of photo- and secondary electrons and from charging of the spacecraft body. We aim to examine these effects by use of numerical simulations in the context of electron measurements acquired by the Electron Analyser System (SWA-EAS) on board the Solar Orbiter mission.}
   {We employed the Spacecraft Plasma Interaction Software to model the interaction of the Solar Orbiter spacecraft with solar  wind plasma. In the model, we implemented a virtual detector to simulate the measured electron energy spectra as observed in situ by the SWA-EAS experiment. For comparison with the real SWA-EAS data, numerical simulations were set according to the measured plasma conditions at 0.3~AU. From the simulation results, we derived the electron energy spectra as detected by the virtual SWA-EAS experiment for different electron populations and compared these with both the initial plasma conditions and the corresponding real SWA-EAS data samples.}  
   {We found qualitative agreement between the simulated and real data observed in situ by the SWA-EAS detector. Contrary to other space missions, the contamination by cold electrons emitted from the spacecraft is seen well above the spacecraft potential energy threshold. A detailed analysis of the simulated electron energy spectra demonstrates that contamination above the threshold is a result of cold electron fluxes emitted from distant spacecraft surfaces. The relative position of the break in the simulated spectrum with respect to the spacecraft potential slightly deviates from that in the real observations. This may indicate that the real potential of the SWA-EAS detector with respect to ambient plasma differs from the spacecraft potential value measured on board. The overall contamination is shown to be composed of emissions from a number of different sources and their relative contribution varies with the ambient plasma conditions.}
   {}


   \maketitle
   \nolinenumbers

\section{Introduction}

Any in situ space instrument for ambient plasma, or, in particular, electron measurements, has to deal with cold electrons emitted from the surface of the spacecraft structure or the surface of the detector itself \citep[e.g.][]{Hutchinson2002}. These emissions are caused either by the absorption of photons, producing the so-called photoelectron emissions, or by impacting (energetic) ambient plasma particles, producing secondary electron emissions. The spacecraft-emitted electrons significantly alter the local plasma properties around the spacecraft and also contribute to the charging of the spacecraft body. The changes in the
local plasma environment and the spacecraft charging in turn disturb the ambient plasma diagnostics. These effects then have to be carefully addressed when the unperturbed plasma properties are derived from the raw electron measurements.

Following the classical theory of \cite{langmuir1924} and \cite{mottsmith1926}, any spacecraft immersed in a plasma environment absorbs fluxes of ambient electrons and ions, while fluxes of photoelectrons and secondary electrons are emitted from its surface simultaneously. The currents associated with these collected and emitted charged fluxes cause electrical charging of the spacecraft structure with respect to the ambient plasma background. An equilibrium is reached at a specific potential of the spacecraft ($\Phi_{SC}$), called the floating spacecraft potential, at which all the charged fluxes are balanced and the total electrical net current to the surface is zero. The final spacecraft potential is thus affected by the ambient plasma properties, solar radiation, and also by specific design aspects of the spacecraft itself \citep{grard1973,garrett1981,whipple1981,pedersen1995,scudder2000}. 

Assuming the UV spectrum to be constant at a certain radial distance from the Sun, the spacecraft potential will decrease with increasing ambient plasma density and temperature, and its actual value will further depend mainly on the ratio of surface areas exposed to and shadowed from the solar radiation. Under typical solar wind conditions, spacecraft are predominantly observed to charge positively, as photoelectron emission usually dominates the collected ambient electron current \citep[e.g.,][]{whipple1981,salem2001}. However, negative charging may occur when the ambient electron current exceeds the photoelectron emission, for example under extreme solar wind densities or temperatures. Negative spacecraft potentials have indeed been reported in the inner heliosphere, based on both in situ observations and numerical simulations \citep{Isensee1981,Guillemant2012AnGeo..30.1075G,Diaz2021,Johansson2020,Fraenz2023}. Mostly positive spacecraft potentials, with occasional negative excursions, were also recently consistently reported for Solar Orbiter \citep{stverak2025}.

\begin{figure*}
    \centering
    \includegraphics[width=1\linewidth]{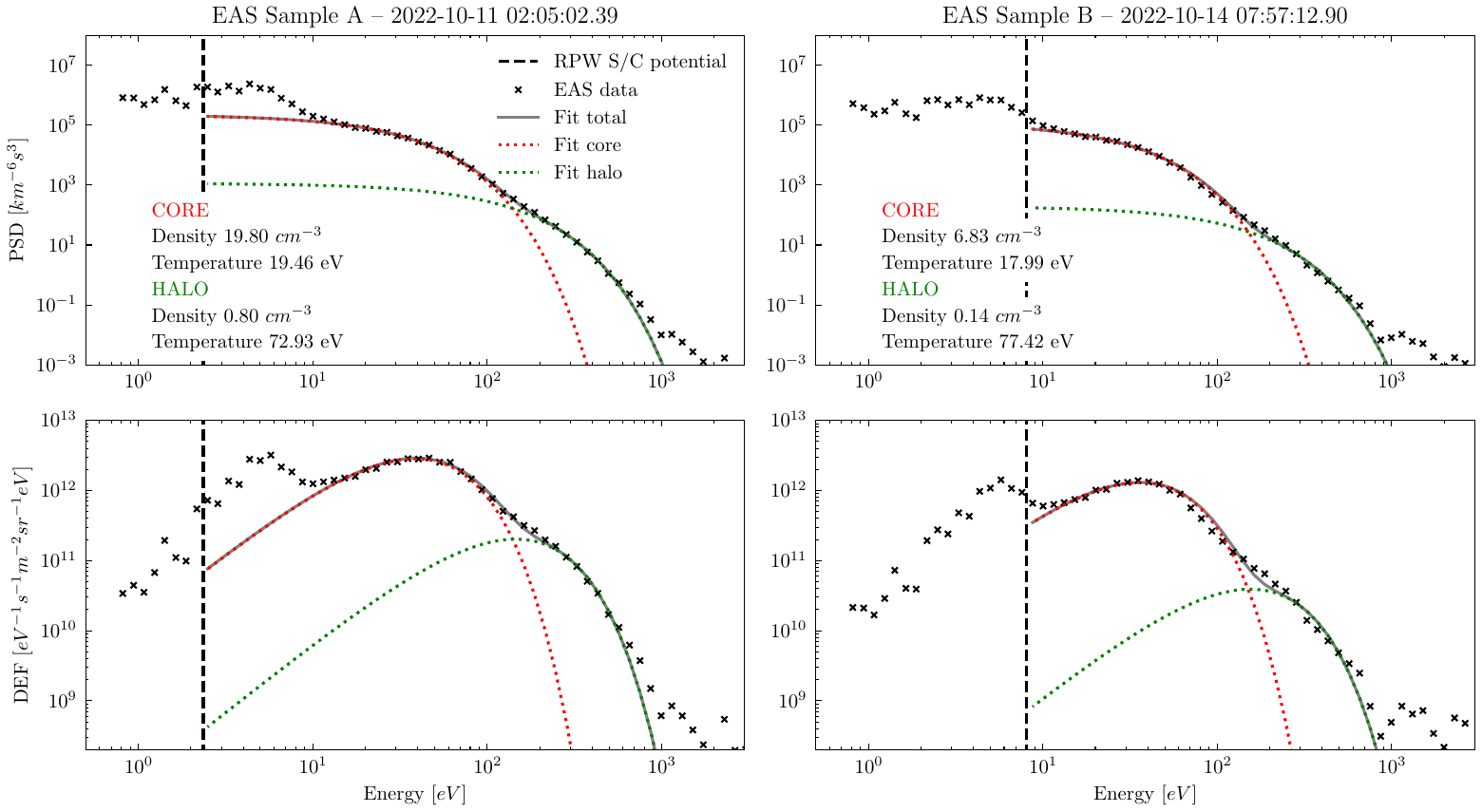}
    \caption{Electron phase space densities (upper panels) and differential energy flux (lower panels) as measured by SWA-EAS are shown (black crosses) as a function of the energy for the selected samples A (left) and B (right). Measured data are over-plotted by a fit with a simple model (gray line) composed from a sum of two Maxwellian distributions for the core (red) and halo (green) ambient electron populations. Displayed fitted plasma parameters are corrected to the spacecraft potential energy measured by RPW (black dashed vertical line).}
    \label{fig:easdata}
\end{figure*}
\begin{table*}
\caption{Basic plasma parameters from Solar Orbiter measurements for selected samples A and B.}
\centering
\begin{tabular}{lcc}
\hline
~ & SAMPLE A & SAMPLE B \\
\hline \hline
Time & 2022-10-11 02:05:02.4 & 2022-10-14 07:57:12.9 \\
Heliospheric distance & 0.2953 AU & 0.2948 AU\\
Electron core/halo density (EAS - model fit) &  19.80 / 0.80  cm$^{-3}$ & 6.83 / 0.14 cm$^{-3}$ \\
Electron core/halo temperature (EAS - model fit) &  19.46 / 72.93  eV & 17.99 / 77.42 eV \\
Electron density (RPW) &  30.20  cm$^{-3}$ & 4.26 cm$^{-3}$ \\
Spacecraft potential (RPW) &  2.33 V & 8.06 V \\
Proton bulk speed\tablefootmark{a} ($v_{x}$,$v_{y}$,$v_{z}$) (PAS) 
    & (-558.1, 31.7, -51.8) km/s
    & (-592.1, 74.4, -29.4) km/s \\
Proton density (PAS) & 66.85\,cm$^{-3}$& 28.45\,cm$^{-3}$\\
Magnetic field strength (MAG) & 46\,nT& 42\,nT\\
\hline \\
\end{tabular}
\tablefoot{
\tablefoottext{a}{The proton bulk speed is given in the Spacecraft Reference Frame (SRF).}
}
\label{tab:solo_parameter}
\end{table*}

Precise knowledge of the spacecraft potential is crucial for any space plasma experiment in order to correctly interpret raw
measurements, in particular, to remove the contamination of the measurements induced by the spacecraft electrons and to correct the energies of ambient electron fluxes. In the simplified scalar approach, the spacecraft potential is assumed to affect only the kinetic energy of the detected particles. All measured energies are then corrected by subtracting the potential energy of the spacecraft $e\Phi_{SC}$ \citep[e.g.][]{genot2004,lavraud2016}. Assuming also changes in the direction of the individual particle trajectories around the positively charged spacecraft body, improved models and methods have to be applied when deriving the unperturbed ambient plasma properties \citep[e.g.][]{scime1994,bouhram2002,hamelin2002,pulupa2014}, particularly in the case of any directional quantities such as the bulk velocity, heat flux, or any anisotropies.

For a positive spacecraft potential, ambient electron fluxes are attracted and accelerated towards the charged spacecraft surfaces. Therefore, all ambient electrons are measured by a detector above an energy of $e\Phi_{SC}$. On the other hand, electrons emitted by the spacecraft at energies below $e\Phi_{SC}$ are trapped by the spacecraft potential and return to the surface. Electrons emitted at energies above the potential threshold $e\Phi_{SC}$ can escape from the surface. Therefore, in principle, any detector should measure the spacecraft electron fluxes at energies $e\Phi_{SC}$ or below. An energy break in the measured energy spectrum of the detected electrons is then expected to exist at $e\Phi_{SC}$, separating the (cold) spacecraft-emitted electron population below $e\Phi_{SC}$ from the (hotter) ambient electron population above $e\Phi_{SC}$. In fact, such a break is often observed in the measured electron spectra and can in turn be used as a possible proxy estimate of the spacecraft potential itself \citep{phillips1993,lewis2008,lavraud2016,wilson2023}. However, most modern plasma missions are capable of directly measuring the spacecraft potential with electric probes or antennas \citep{pedersen1995,scudder2000,pedersen2008,lindqvist2016}.

\begin{figure*}
    \centering
    \includegraphics[width=.9\linewidth]{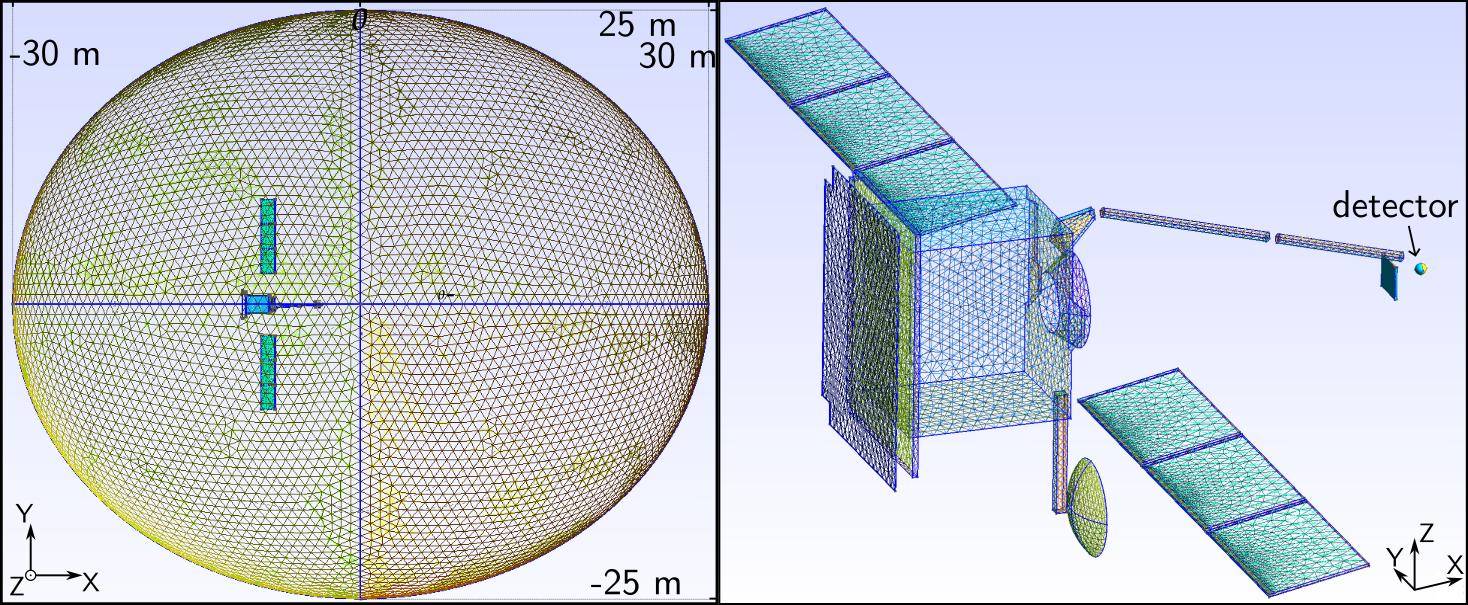} 
    \caption{The computational mesh used in the simulation model. The left panel shows the whole computational volume (view from top) comprised in an ellipsoid with 30\,m and 25\,m long semi-axes along the main X and Y axes (and 20\,m semi-axes along Z). The right panel shows the surface mesh model of the Solar Orbiter with solar panels rotated at an angle of 79$^{\circ}$ to Sun normal, reflecting the actual geometry configuration at 0.3\,AU. The SWA-EAS detector is modelled as a single sphere at the tip of the payload boom.}
    \label{fig:solo-gmsh-model}
\end{figure*}

On the Solar Orbiter mission \citep{muller2020,zouganelis2020}, the spacecraft potential is provided by a set of DC voltage measurements from three electric antennas attached to the main spacecraft body \citep{khotyaintsev2021}. Measurements of ambient electron fluxes are acquired by two top-hat electrostatic analysers with aperture deflectors mounted at the tip of a payload boom deployed behind the spacecraft body with respect to the incoming stream of solar wind plasma \citep{owen2020}. The contamination of ambient electron measurements on board the Solar Orbiter by the cold electrons emitted from the spacecraft under the effect of the induced spacecraft potential was recently investigated by \citealt{stverak2025}. They observed a break in the electron energy spectra between the spacecraft and ambient electron populations but its location was found to be largely uncorrelated with the measured spacecraft potential values, and significant fluxes of spacecraft-emitted electrons were detected even above
the spacecraft potential threshold. \citealt{stverak2025} proposed an explanation for this discrepancy, based on the detection of spacecraft electrons emitted by surfaces with the same potential, but far from the detector itself. The observed location of the break in the electron energy spectrum was found to be consistent with this scenario, implying a theoretical dependency of the break location on the ambient electron temperature. However, current in situ plasma measurements, where any electrons are not directly separable by their origin, do not allow us to fully confirm this effect. 

In this paper, our objective is to support and further investigate the conclusions of \cite{stverak2025} using numerical modelling and simulations of solar wind plasma interactions with the Solar Orbiter spacecraft. We have followed, and further extended, the previous work of \cite{guillemant2012,guillemant2014,guillemant2017} to provide virtual measurements of different individual electron populations according to the realistic configuration of the Solar Orbiter spacecraft. In particular, we performed two case studies corresponding to real Solar Orbiter measurements, as described in Sect.~\ref{sec:data}. The numerical model and the simulation results of the two case studies are presented in Sect.~\ref{sec:model} and Sect.~\ref{sec:results}, respectively. The simulation results are discussed in the context of the real Solar Orbiter observations in Sect.~\ref{sec:dsc}, and the main findings are summarised in Sect.~\ref{sec:con}.

\section{Measurements}
\label{sec:data}

All data used in our study are available via the public Solar Orbiter archive\footnote{ESA Solar Orbiter Archive is available at \url{http://soar.esac.esa.int}} (SOAR) provided by the European Space Agency (ESA). Electron measurements on board the Solar Orbiter spacecraft are provided by the Electron Analyser System (EAS), which is part of the Solar Wind Analyser (SWA) suite of plasma instruments \citep{owen2020}. The SWA-EAS instrument is designed to measure full 3D electron velocity distribution functions (VDFs) using two top-hat electrostatic analysers, EAS1 and EAS2, mounted on the main spacecraft boom, which extends anti-sunward into the shadow behind the spacecraft body. The combined field of view of the two detectors covers almost the full $4\pi$~sr, although a small part of the sky is obstructed by the mechanical structures of the spacecraft and the instrument itself (see Figure~3 in \cite{owen2020}). The measured energy ranges from $<1$\,eV up to~5\,keV in~64 logarithmically spaced steps, with an energy bandwidth of $\Delta E / E \approx 13\%$. In the present work, for the initial setup and subsequent comparison with numerical simulations, we used the Level~2 data product of phase space densities (PSD) and differential energy flux (DEF) measured in Normal Mode 3D (NM3D), see \cite{owen2020}. 

For our study, we selected two data samples (sample~A acquired on 2022-10-11 at 2:05 UTC, and sample~B acquired on 2022-10-14 at 7:57 UTC) representative for SWA-EAS measurements taken during the second perihelion of Solar Orbiter at about 0.3~AU from the Sun. The SWA-EAS 3D distribution function samples were further processed, following \citet{stverak2025}, into reduced 1D energy spectra averaged over all instrument azimuths and elevations according to (\ref{eq:fov}). The two selected samples, converted into the 1D energy distribution, are plotted in Fig.~\ref{fig:easdata}. The upper panels show the energy distribution of phase space density while the lower panels plot the energy distribution of the differential energy flux. The latter quantity allows easier visual separation of the different electron populations, namely the spacecraft emitted electrons at lowest energy range (below 10\,eV in our case), core electrons in the thermal energy range (up to about 100\,eV), and halo electrons at suprathermal energies (above 100\,eV). The spacecraft potential as measured by RPW is plotted for both samples with the vertical dashed line.

The two samples analyzed in this work were intentionally selected  as illustrative case studies rather than as representative of the average plasma conditions. Both samples were taken near perihelion, i.e., within the Solar Orbiter mission target environment, but were chosen to bracket two contrasting charging regimes observed by SWA-EAS and RPW (\cite{stverak2025}): (sample~A) a low-potential case for which the energy break in the measured electron spectrum between the spacecraft-emitted and core solar wind electron populations occurs well above the RPW spacecraft potential, and (sample~B) a higher-potential case for which the break occurs close to the RPW potential. The goal of this selection is to test the model performance under two distinct observational configurations and to disentangle the contribution of spacecraft-emitted electrons, rather than to infer typical or mean conditions at 0.3~AU.

The spacecraft potential is measured on board the Solar Orbiter by the Radio and Plasma Waves (RPW) experiment \citep{maksimovic2020}. The RPW instrument is primarily designed to measure the local magnetic and electric fields and their fluctuations as electromagnetic wave spectra by the use of a three-axis search coil magnetometer and a set of three electric monopole antennas. The spacecraft potential is estimated by measuring the DC voltages on the three antennas with respect to the spacecraft body \citep{khotyaintsev2021,steinvall2020,maksimovic2021}. Calibrated values of the spacecraft potential with respect to the plasma used in our study are available as the RPW Level~3 BIAS SCPOT data products. The RPW Level~3 data further provide the electron density as the BIAS DEN data product derived from the measured spacecraft potential and its cross-calibration to the measured plasma frequency. 

The electron density and other electron moments from SWA-EAS measurements are not (yet) publicly available in SOAR. For comparison with RPW measurements, we estimate the electron density and temperature by fitting the measured phase space density spectra in the thermal energy range (20-70\,eV) for the core and in the suprathermal energy range (200-600\,eV) for the halo electron population. The estimated electron densities were corrected for the effect of the spacecraft potential as measured by RPW by use of the Boltzmann factor $\exp \left( -e\Phi/kT\right)$. The fitting model uses two isotropic non-drifting Maxwellian distributions defined by (\ref{eq:maxwellPRF}) as shown in Fig.~\ref{fig:easdata} by red and green dotted lines for the core and halo electron population, respectively. 

The electron (or solar wind) bulk speed is difficult to be derived from the SWA-EAS measurements. We used the proton bulk speed from the on-ground moments of the ion 3D VDFs acquired by the Proton and Alpha Sensor (PAS), which is another part of the SWA experiment, as reported in \cite{owen2020}. The on-ground SWA-PAS proton moments are available in SOAR as Level~2 GRND MOM data products. For a complete description of the plasma background environment, including the relevant characteristic spatial and temporal electron scales, we used measurements of the magnetic field acquired by the Solar Orbiter magnetometer (MAG, see \cite{horbury2020}). All measured data used in the present study are given in the Spacecraft Reference Frame\footnote{ESA Solar Orbiter SPICE kernels are available at \url{https://spiftp.esac.esa.int/data/SPICE/SOLAR-ORBITER/}} (SRF). All relevant plasma parameters from the in situ Solar Orbiter instruments for the two selected SWA-EAS samples are summarized in Tab.~\ref{tab:solo_parameter}. The difference in the derived plasma electron densities between EAS and RPW likely reflects remaining uncertainties in the cross-calibration of the available Solar Orbiter data products. The RPW densities, which are internally calibrated using the plasma frequency measurements and are therefore considered to provide a more reliable estimate of the ambient plasma density, are used as the reference values in this study.

\section{Numerical model}
\label{sec:model}
In order to model the interaction of the Solar Orbiter spacecraft with the ambient solar wind plasma, and thus to identify and distinguish possible sources of the spacecraft emitted electrons, we used the Spacecraft Plasma Interaction Software (SPIS, version 6.2.4.),see \citep{Hess2024_spisv6,Sarrailh2014}. SPIS is a free open source software tool for matter-plasma interaction modelling which can use either hybrid or full particle-in-cell (PIC) approach to describe the behaviour of the plasma environment. In particular, the software enables modelling the spacecraft geometry, defining the spacecraft surface parameters, setting the ambient plasma properties, and applying number of parameters for the interaction between the plasma and the spacecraft itself. 

\begin{table}
\caption{SPIS model geometry groups, materials, and nodes.}
\centering
\begin{tabular}{ccc}
\hline
group & material & node\\
\hline \hline
spacecraft & Black Kapton & 0 \\
Sun shield & Black Kapton & 1\\
solar array front &  solar cell & 3,4\\
solar array back &  Black Kapton & 3,4 \\
boom & Black Kapton & 0\\
eas sensor & steel & 0 \\
eas baffle & Black Kapton & 0\\
HGA & electrodag black paint & 2 \\
HGA boom & electrodag black paint & 2 \\
\hline \\
\end{tabular}
\label{tab:simulation_groups}
\end{table}

Our model is based on the real, but simplified, Solar Orbiter geometry, adapted to represent the main structural components: spacecraft body, Sun shield, solar panels, high-gain antenna with its boom, main payload boom with protective SWA-EAS baffle, and SWA-EAS sensor itself. The model and its setup are adopted, according to the geometry and material selection, from \citet{guillemant2017} by adding additional modifications to better reflect the real spacecraft parameters and materials. The geometric surface mesh of the model and of the simulation volume boundary is shown in Figure~\ref{fig:solo-gmsh-model}. The SWA-EAS sensor itself is represented as a simple spherical surface, with a diameter of 12\,cm, used as a support for individual virtual SPIS particle detectors used to monitor all relevant plasma populations. The sphere is placed at the end of the spacecraft boom behind the baffle (see the position in the right panel of Fig.~\ref{fig:solo-gmsh-model}), reflecting the real position of the SWA-EAS sensors. The model therefore differs from the real geometry of the SWA-EAS instrument, with its two separated cylindrical structures of the top-hat analysers EAS~1 and~2 mounted on the SWA-EAS electronic box (cf. Fig.~4 in \cite{owen2020}), but is still well representative in terms of the real position of the sensor with respect to the spacecraft. 

\begin{table*}
\caption{Initial parameters for SPIS simulations and the relevant characteristic electron plasma scales.}
\centering
\begin{tabular}{lcc}
\hline \hline
Parameter & RUN A & RUN B \\
\hline 
Heliospheric distance & \multicolumn{2}{c}{0.295 AU}\\
Electron/proton density &  30  cm$^{-3}$ & 4.3 cm$^{-3}$ \\
Electron/proton temperature &  19.5 eV & 18 eV \\
SW bulk speed ($v_{x}$,$v_{y}$,$v_{z}$) 
    & (558, -31.7, -51.8) km/s
    & (592, -74.4, -29.4) km/s \\
Photoelectron temperature & \multicolumn{2}{c}{3 eV}\\
Secondary electron temperature & \multicolumn{2}{c}{2 eV}\\
Average macroparticles per cell & \multicolumn{2}{c}{200}\\
Simulation time & \multicolumn{2}{c}{1.5 s}\\
Computational box dimensions (ellipsoid) & \multicolumn{2}{c}{60x50x40 m$^{3}$} \\ 
Spacecraft capacitance & \multicolumn{2}{c}{1 $\mu$F}\\
\hline \hline
Characteristic scales & ~ & ~ \\
\hline
Debye length ($\lambda_{D}$) & 6.0\,m & 15.2\,m \\ 
Electron gyroradius ($r_{c,e}$) & 229\,m & 251\,m \\ 
Electron plasma frequency ($f_{p,e}$) & 49.2\,kHz & 18.6\,kHz \\ 
\hline \\
\end{tabular}
\label{tab:input_parameter}
\end{table*}

The total computational (plasma) volume is a rotational ellipsoid with semi-axes of 30\,m, 25\,m, and 20\,m along the X, Y, and Z direction, respectively. The orientation of the X and Y axes is reversed compared to SRF. The spacecraft itself is moved by 9~m in the -X direction (see Fig.~\ref{fig:solo-gmsh-model}) to well resolve the plasma wake along +X axis inside the computational volume. SPIS allows for a variable mesh resolution. In our case, the mesh resolution at the external boundary is reduced considerably compared to the resolution close to the spacecraft to save computational resources. Still, the spacecraft is meshed with a grid size small enough to correctly resolve the Debye length and particle trajectories close to the plasma detector. For the spacecraft body, the resolution varies between 8~and 15\,cm and at the SWA-EAS sensor the grid size is set to 4\,cm. At the external boundary of the computational volume, the grid size is set to 1\,m and is further reduced to 0.5\,m towards the spacecraft body. In the whole computation domain the grid resolution thus decreases with distance from the spacecraft, to keep reasonable simulation times, but still capable of resolving expected local potential field or density gradients. 

The geometry of the spacecraft is divided into several structural groups with dedicated surface material properties and corresponding electrical nodes, as given in Tab.~\ref{tab:simulation_groups}. Most of the spacecraft surface area is set to behave as a black kapton foil, and a conductive graphite antistatic black paint is used for the high-gain antenna. The SWA-EAS sensor is modelled as a steel surface. The electric circuit, representing the individual structural groups of the spacecraft, has a star configuration with node~0 (spacecraft) as a central node and all other surrounding nodes connected by an effective resistance R. While the ideal conductive limit corresponds to R\,=\,0, we adopt a finite value R = 1\,M$\Omega$ to represent imperfect electrical coupling and contact resistances between different spacecraft structures. This parameter should be understood as an effective modelling quantity rather than a precisely known hardware property. The total spacecraft capacitance is set to 1\,$\mu$F as an upper limit value for the expected capacitance range for a spacecraft of Solar Orbiter’s size. While the capacitance does not affect the final equilibrium spacecraft potential, which is set by current balance, it controls the temporal response of the charging process in the simulation. The adopted value of 1\,$\mu$F is chosen to reduce potential fluctuations due to the limited particle statistics in the simulation, while remaining small enough to ensure that a steady-state potential is reached within a fraction of the total simulation time.

\begin{figure*}
    \centering
    \includegraphics[width=1\linewidth]{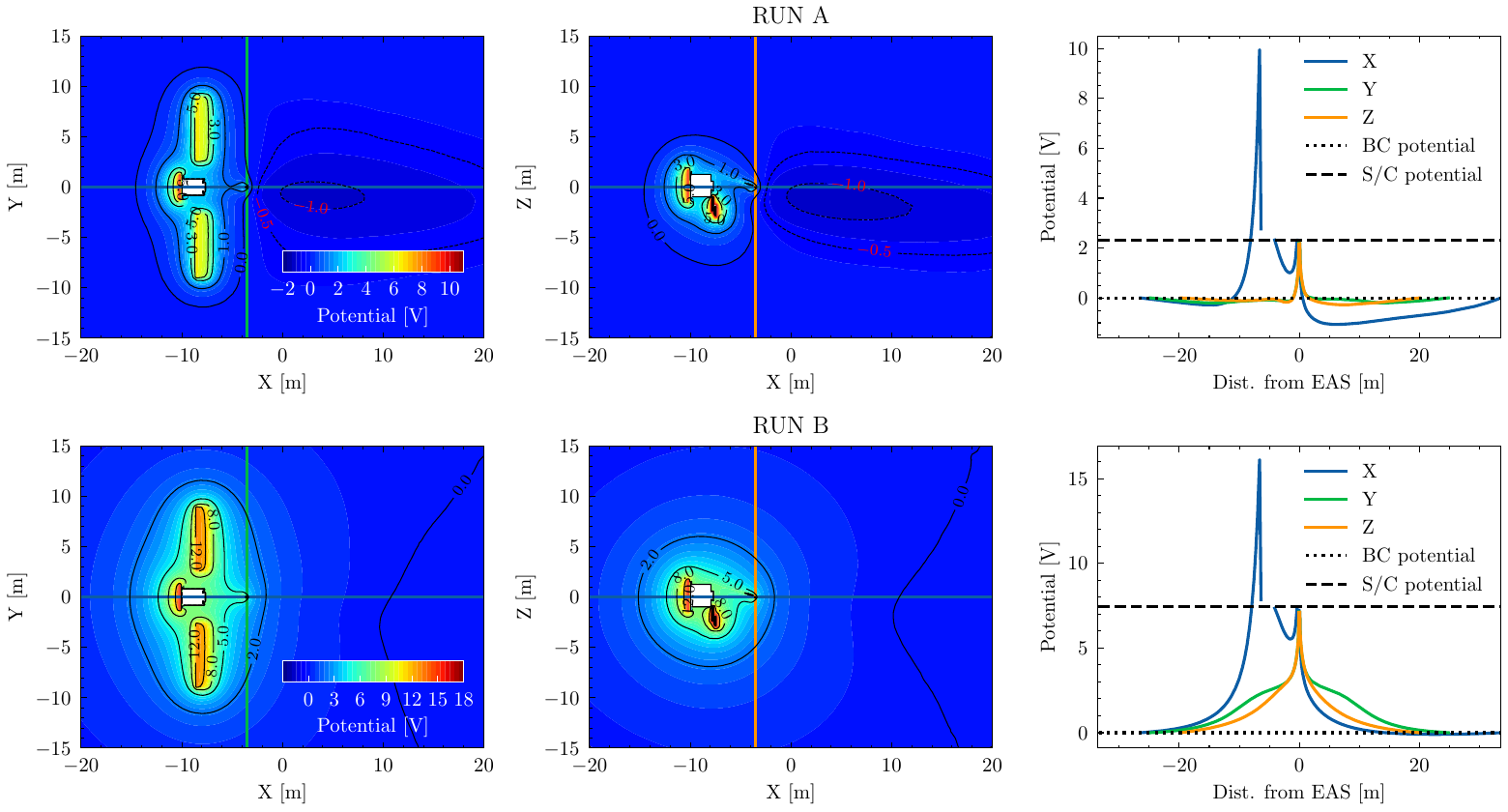}
    \caption{Final structure of the potential around the spacecraft body is shown for simulation run~A (top row) and run~B (bottom row) at time t=1.5\,s. The left and middle column show the 2D cuts in the XY and XZ planes, respectively. The right column shows potential profiles along X (blue), Y (green), and Z (orange) axis as a function of the distance from the virtual SWA-EAS detector. The dashed line in the right panel displays the final surface spacecraft potential and the dotted line line shows the background (zero) plasma potential for reference. }
    \label{fig:potentials}
\end{figure*}

The initial ambient plasma conditions were set according to the selected SWA-EAS sample~A and sample~B for the simulation run~A and run~B, respectively. In particular, we set the ambient plasma density according to the electron density measured by RPW, ambient plasma temperature was set to the core electron temperature from fits of SWA-EAS data samples, and for plasma bulk speed we used the proton measurements performed by SWA-PAS.
The effective temperature of the photo and secondary electron emissions was set to 3 and 2\,eV, respectively, according to \cite{guillemant2017}.
The general overview of the initial input parameters for both simulations is given in Tab.~\ref{tab:input_parameter}, cf. the original values as measured by Solar Orbiter given in Tab.~\ref{tab:solo_parameter}. Table~\ref{tab:input_parameter} further shows the relevant characteristic plasma scales for the initial background plasma environment. 

The size of the computational volume, being limited by available computational resources, was set to respect the Debye length derived for initial ambient plasma conditions, see actual values of $\lambda_{D}$ in Tab.~\ref{tab:input_parameter}. The minimum distance from any surface of the spacecraft to the boundary of the simulation box is thus always at least $\gtrsim \lambda_{D}$. The physical simulation time of 1.5\,s was set to be significantly greater than both the inverse of the electron plasma frequency and the plasma transition time through the simulation box given by the solar wind bulk speed. 
SPIS is implemented as an electrostatic plasma code and thus computes the electric field (or potential) by solving the Poisson equation. For the spacecraft surfaces, the boundary condition is set by default to Dirichlet type so that the potential is constant over the surface of the individual nodes of the spacecraft structure. For the outer surface boundary of the simulation domain, we also used the Dirichlet condition with the boundary value of the potential set to zero to reflect the ambient plasma conditions.
We further neglected the magnetic field in our model (B is set to 0) as the electron gyroradius of the thermal ambient plasma ($\gtrsim 200$ km/s) is much greater than the size of the spacecraft and even than the size of the simulation box. The magnetic field can also be neglected in terms of $v \times B$ as for the selected solar wind plasma conditions the related convective electric field is below 20~mV/m and thus rather small compared to the typical electric field induced by the charged spacecraft. 

The individual plasma populations simulated in the model were ambient ions/protons (AI), ambient electrons (AE), photoelectrons (PE), secondary electrons from electron impact (SE), and secondary electrons from ion impact (SI). Both the ambient proton and electron populations were modelled as drifting isotropic Maxwellian distributions using the full PIC approach. We therefore considered the thermal part of the solar wind plasma only while any suprathermal tails (halo) or beams with relative drifts (strahl) were neglected as less relevant for the present study in order to keep the model simple and save the limited computational resources. Default SPIS properties were used for the selected surface materials to define the production rates of photo and secondary electron populations. An amplification parameter of~11.5 was applied to the default UV flux referenced in SPIS to 1~AU, reflecting the actual heliocentric distance 0.295~AU of the selected data samples.

The SWA-EAS instrument was implemented as a virtual SPIS plasma detector defined to cover the entire spherical surface of its structural support. The sensor was activated to register the incoming particle fluxes of all individual electron species separately with a full 4$\pi$ field of view. An ideal model response of such detector is evaluated in Appendix~\ref{sec:appdet}.
The so called particle list data products, produced by the individual SPIS virtual SWA-EAS detectors, were consequently used to derive the electron energy distribution function for all electron populations as given by (\ref{eq:fspis}). The SPIS particle list, representing a statistical sample of detected electrons, includes both the initial (emission) and final (detection) particle positions and velocities and thus provides the possibility to derive not only the total particle energy distributions, but also partial distributions for individual particle fluxes from different source locations of their emission. We produced particle lists for all electron plasma species at a cadency of 20\,Hz and used the last 10 samples of the simulation run time summed and averaged for the later analysis of the simulated electron distribution functions.

\section{Results}
\label{sec:results}

\begin{figure*}
    \centering
    \includegraphics[width=1\linewidth]{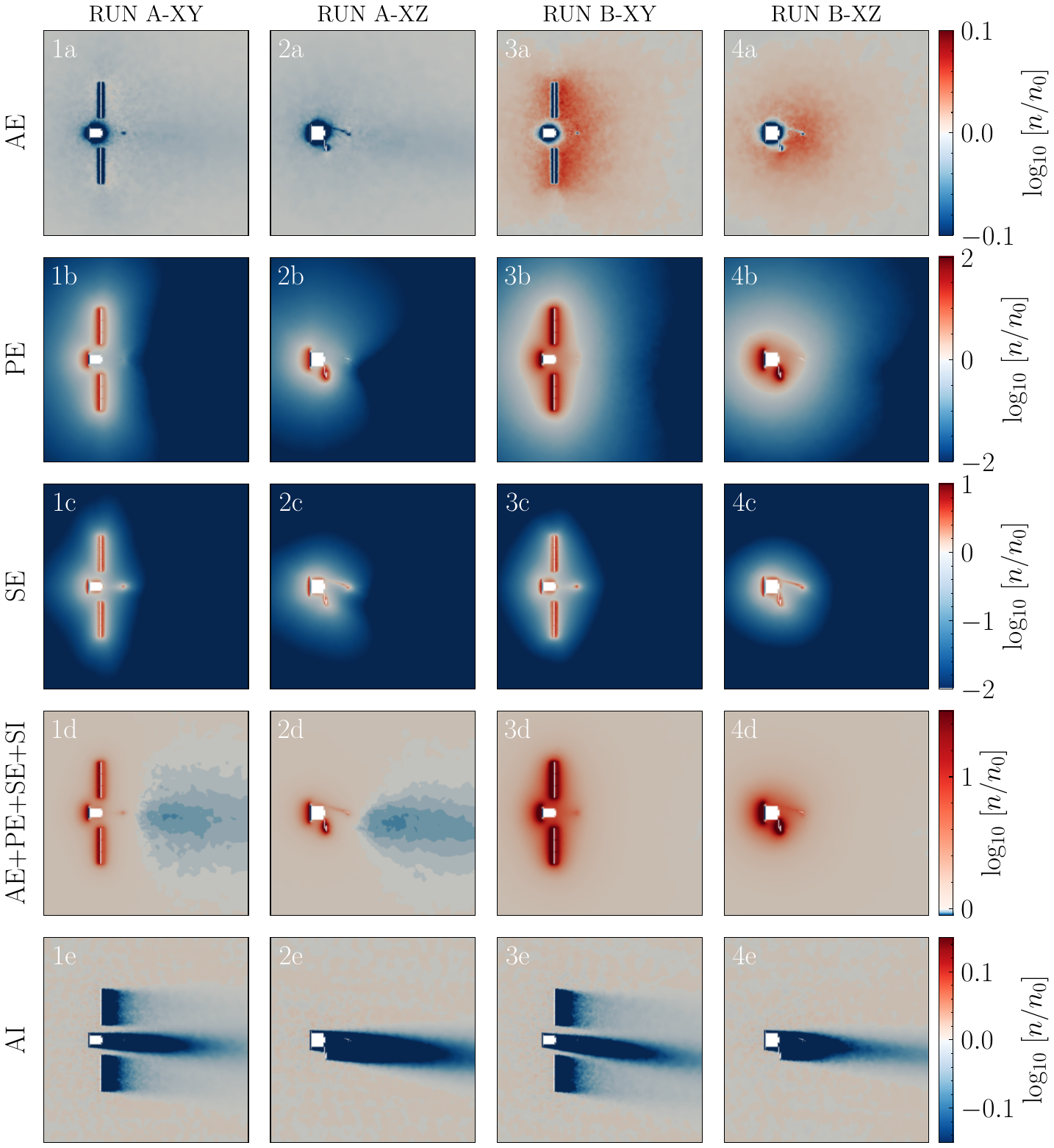}
    \caption{The electron an ion densities at the final simulation time t=1.5\,s are show for run~A (columns 1 and 2) and run~B (columns 3 and 4) as 2D slices in the XY and XZ plane: row 1a-4a for ambient electrons (AE), row 1b-4b for photoelectrons (PE), row 1c-4c for secondary electrons from electron impacts (SE), row 1d-4d total electron density (AE+PE+SE+SI), and row 1e-4e for ambient ion density (AI). All densities are normalized to the initial ambient plasma density n$_{0}$ and plotted in logarithmic scale showing relative increase in densities in red and decrease in densities in blue colours.}
    \label{fig:densities}
\end{figure*}

We performed the SPIS simulations according to two different initial background plasma conditions and using a simplified geometrical model of the Solar Orbiter spacecraft as described in sections \ref{sec:data} and \ref{sec:model}, respectively. The two simulation studies correspond to the case of high plasma density and thus low spacecraft potential (run~A) and to the opposite case with rather low plasma density and high spacecraft potential (run~B). In both simulations, the spacecraft charging process reached steady state well after a simulation time of approximately 0.3\,s. For the rest of the simulation the surface potentials were almost constant with a relative standard deviation below 1\%. For results presented in this section we used the final state of the simulations at the time of 1.5\,s to plot the potential and plasma properties around the spacecraft body, and individual electron distribution functions averaged over a time interval of $\left\langle 1.0, 1.5 \right\rangle$\,s.

The resulting mean equilibrium surface potentials on individual electric nodes are listed in Tab.~\ref{tab:potentials}. For node~0, that is, the main spacecraft body with the payload boom and including the virtual SWA-EAS detector, the equilibrium potential is reached at 2.31\,V and 7.37\,V for run~A and run~B, respectively, cf. the spacecraft potential values of 2.33\,V and 8.06\,V measured by RPW for sample~A and sample~B, respectively, as given in Tab.~\ref{tab:solo_parameter}. Significantly higher potentials are reached on the other sun-lit surfaces of the remaining nodes, isolated in our model from the spacecraft ground with a resistance of 1\,M$\Omega$. The complex spatial structure of the potential around the spacecraft body is shown in Fig.~\ref{fig:potentials} by plotting the potential maps as 2D slices in the XY and XZ planes and 1D potential profiles along lines parallel to all three main axes X, Y, and Z centred at the position of the SWA-EAS detector. The perturbation of the ambient plasma potential induced by the positively charged surfaces of the spacecraft is observed even in a distance of about 5\,m for run~A and further away for run~B, being consistent with the local plasma Debye length (cf. Tab.~\ref{tab:input_parameter}). In the case of run~A, a negative potential sheath is created around the spacecraft body, with minimum values in the wake behind the spacecraft, while the potential remains positive within the simulation domain in the case of run~B. The negative potential sheath in run~A results in a non-monotonic potential profile from the detector to the boundary, mainly along the +X axis, as indicated in the upper left panel of Fig.~\ref{fig:potentials}. In run~B, the potential drops monotonically from the positive surface values to the zero potential given by the boundary condition for the ambient plasma environment. The non-monotonic potential profile from the detector along the -X axis is due to the crossing of the spacecraft body and the front heat shield.

\begin{table}
\caption{Final surface potentials for electrical nodes.}
\centering
\begin{tabular}{clcc}
\hline
node ID & Description & RUN A & RUN B\\
\hline \hline
0 & SC & 2.31\,V & 7.37\,V \\
1 & Sun shield & 9.91\,V & 16.12\,V \\
2 & HGA & 12.06\,V & 17.95\,V \\
3 & Solar Arrays & 7.47\,V & 14.0\,V \\
\hline \\
\end{tabular}
\tablefoot{Values are averaged over the simulation time interval $\left\langle 1.0, 1.5 \right\rangle$\,s.}
\label{tab:potentials}
\end{table}

The results of both simulations in terms of the local densities of electrons and ions around the spacecraft are presented in Fig.~\ref{fig:densities}. The plots show slices in the XY plane (z=0) and XZ plane (y=0) for all individual plasma species. Panels 1a-4a show the density of the ambient electron population. A drop in the ambient electron density is observed near the positively charged spacecraft surfaces, where the electron density is dominated by photo and secondary electrons; see panels 1b-4b for photo and 1c-4c for secondary emissions, respectively. The local density of photoelectrons near the spacecraft surface is about 100 times higher than the background plasma density in run~B. The concentrations of secondary electrons are about an order of magnitude lower compared to photoemissions. However, the concentration of secondary electrons near the location of the SWA-EAS detector is similar to (run~B) or even significantly higher (run~A) than the local concentration of photoelectrons. In general, the cold electrons emitted from the spacecraft surfaces significantly distort the ambient plasma around the spacecraft body up to a distance of about 5\,m. The total electron density (AE+PE+SE+SI) is plotted in panels 1d-4d showing a structure consistent with the potential profiles, cf. Fig.~\ref{fig:potentials}. In panels 1e-4e we plot the ambient ion densities. A distinct ion wake is created behind the spacecraft structures along the +X axis by the solar wind flow, which is known to be supersonic compared to the ion thermal velocities. Secondary electrons due to ion impacts are produced on surfaces facing the ram direction to the solar wind flow, and their spatial distribution is very similar to photoelectrons, however, their contribution to the total electron content around the spacecraft is negligible ($\lesssim$~1\%, not shown in Fig.~\ref{fig:densities}).

\begin{figure*}
    \centering
    \includegraphics[width=0.95  \linewidth]{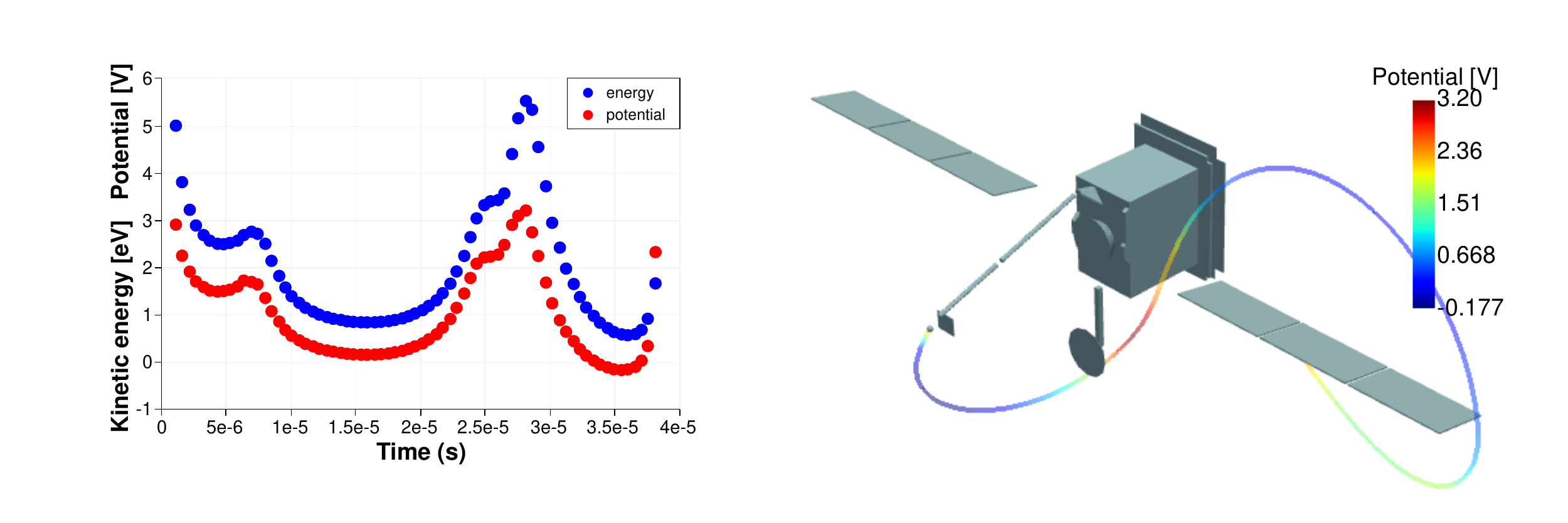}
    \caption{Example of photoelectron trajectory emitted from the solar panel and impacting the surface of the SWA-EAS detector. Change in the electron kinetic energy (blue) and potential (red) along the trajectory is shown in the left panel as a function of the time of flight. The sample trajectory is taken from the simulation run~A.}
    \label{fig:pe-trajectory}
\end{figure*}

The complex structure of the potential (and thus also electric) field created around the spacecraft naturally affects the trajectories of both ambient and spacecraft-emitted electrons with significant deviations, particularly caused to particles with kinetic energies comparable to or below the spacecraft potential energy. The right panel of Fig.~\ref{fig:pe-trajectory} shows a sample trajectory of a single photoelectron emitted on the front side of the solar panel that travels in the spacecraft potential field and finally hits the position of SWA-EAS detector. The left panel shows a plot of the change in the kinetic energy of the particle as a function of time along its trajectory. The electron loses and gains kinetic energy along the path, based on the potential of the nearby spacecraft structures. Moving away from the positively charged solar panel into the ambient background plasma, the electron is first decelerated and reaches an almost zero potential at a time $\approx$~16\,$\mu$s, with a minor partial acceleration around~8\,$\mu$s near the tip of the panel. Subsequently, the electron is attracted and accelerated by the potential of the spacecraft body and later by the potential of the high-gain antenna, reaching the maximum kinetic energy at $\approx$~27\,$\mu$s. Finally, it moves away from the spacecraft into the wake but is reflected by the negative potential well (cf. Fig.~\ref{fig:potentials}) and accelerated towards the SWA-EAS detector. The net change in the particle's kinetic energy is negative in accordance with the difference between the surface potential of the solar panels and the potential of the SWA-EAS (spacecraft) surface, cf. Tab.~\ref{tab:potentials}.

\begin{figure*}
    \centering
    \includegraphics[width=1\linewidth]{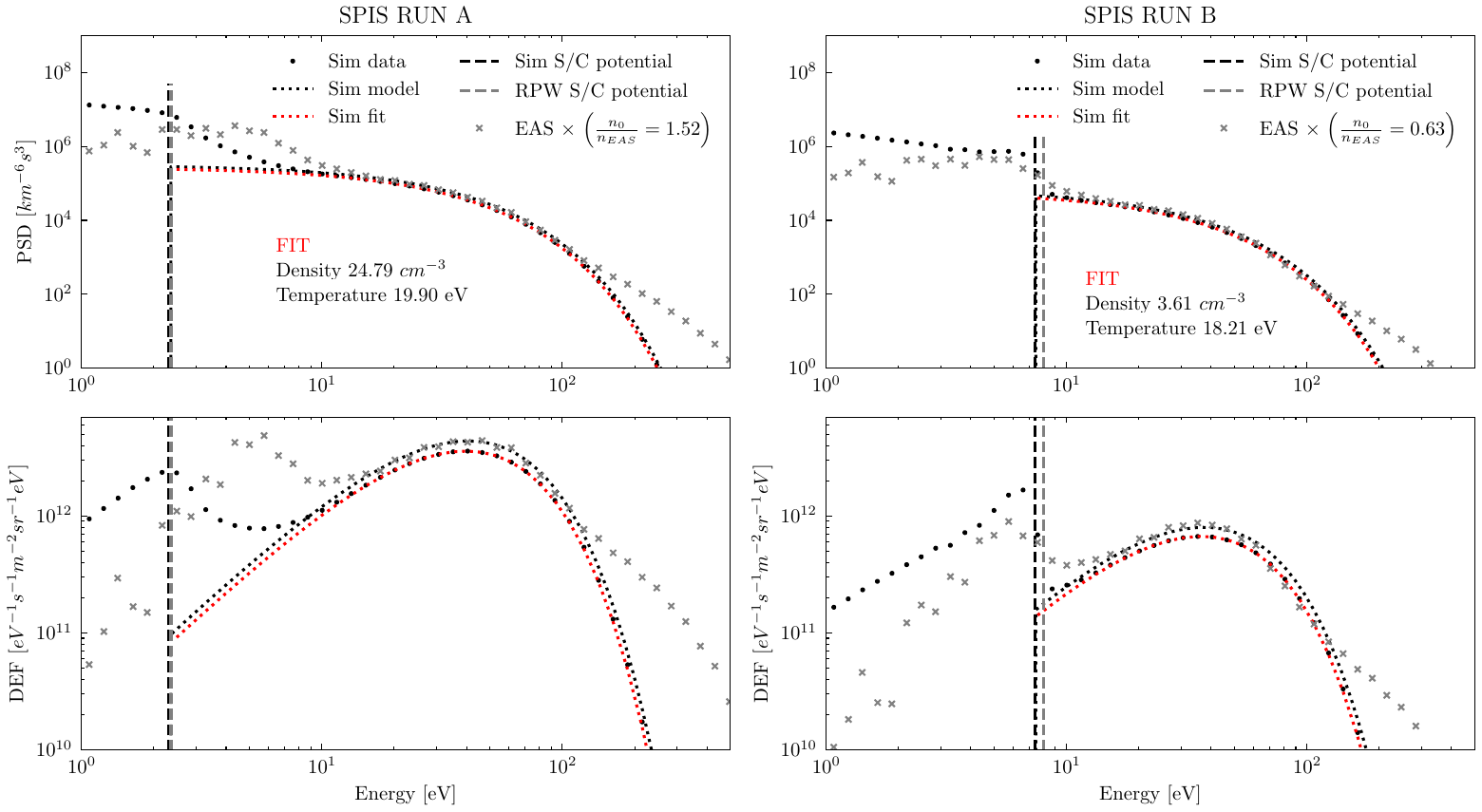}
    \caption{Phase space densities (upper panels) and differential energy flux (lower panels) are compared between the simulation results (black dots) and real SWA-EAS measurements (grey crosses) for both run~A (left) and run~B (right). The measured SWA-EAS data are scaled by the ratio of the initial plasma density ($n_0$) and the core electron density from SWA-EAS ($n_{EAS}$). Model response of the detector to drifting Maxwellian distribution is shown for simulated ambient plasma conditions (black dotted line) and simulated data are fitted by a Maxwellian model (red dotted line). Electron parameters derived from the fit are corrected to the simulated spacecraft potential (black dashed line) and the spacecraft potential as measured by RPW is shown for comparison (grey dashed line).}
    \label{fig:vdfA_v0}
\end{figure*}

The general effect of the spacecraft surface potentials and of the emitted spacecraft electrons on the SWA-EAS measurements was analysed using 1D integrated energy distribution functions, time-averaged over simulation times $\left\langle 1.0, 1.5 \right\rangle$\,s. The 1D energy distribution functions obtained by the virtual SWA-EAS SPIS detector from both simulation runs are shown in Figure~\ref{fig:vdfA_v0}, in the same form as the real data samples (cf. Figure~\ref{fig:easdata}). The simulated distributions (black dots) are directly compared with the corresponding real measurements (grey crosses) from the SWA-EAS data samples. The measured SWA-EAS values are multiplied by a correction factor given by the ratio of the initial ambient plasma density, set in the simulation according to the RPW measurements, and the electron density derived from the real SWA-EAS measurements (cf. Tab.~\ref{tab:solo_parameter} and Tab.~\ref{tab:input_parameter}) to match the expected theoretical response from the simulation for ambient electrons (black dotted line) given by (\ref{eq:fdetector}). The plots further show a comparison of the spacecraft potential measured by RPW (grey dashed vertical line) to the spacecraft surface potential resulting from the simulation (black dashed vertical line). The same fitting procedure and correction to the spacecraft potential as for the measured SWA-EAS samples is performed, but with a single Maxwellian distribution (red dotted line), to the simulated electron energy spectra in the thermal electron energy range (20-70\,eV). The values of electron density and temperature derived from the fit to the virtual SWA-EAS measurements are found to be comparable to the initial unperturbed plasma conditions, cf. Tab.~\ref{tab:input_parameter}. The theoretical Maxwellian profile of ambient electrons is significantly disturbed at low energies ($\lesssim 10$\,eV) for both simulation runs, qualitatively similar to the real SWA-EAS measurements.  

\begin{figure*}
    \centering
    \includegraphics[width=1\linewidth]{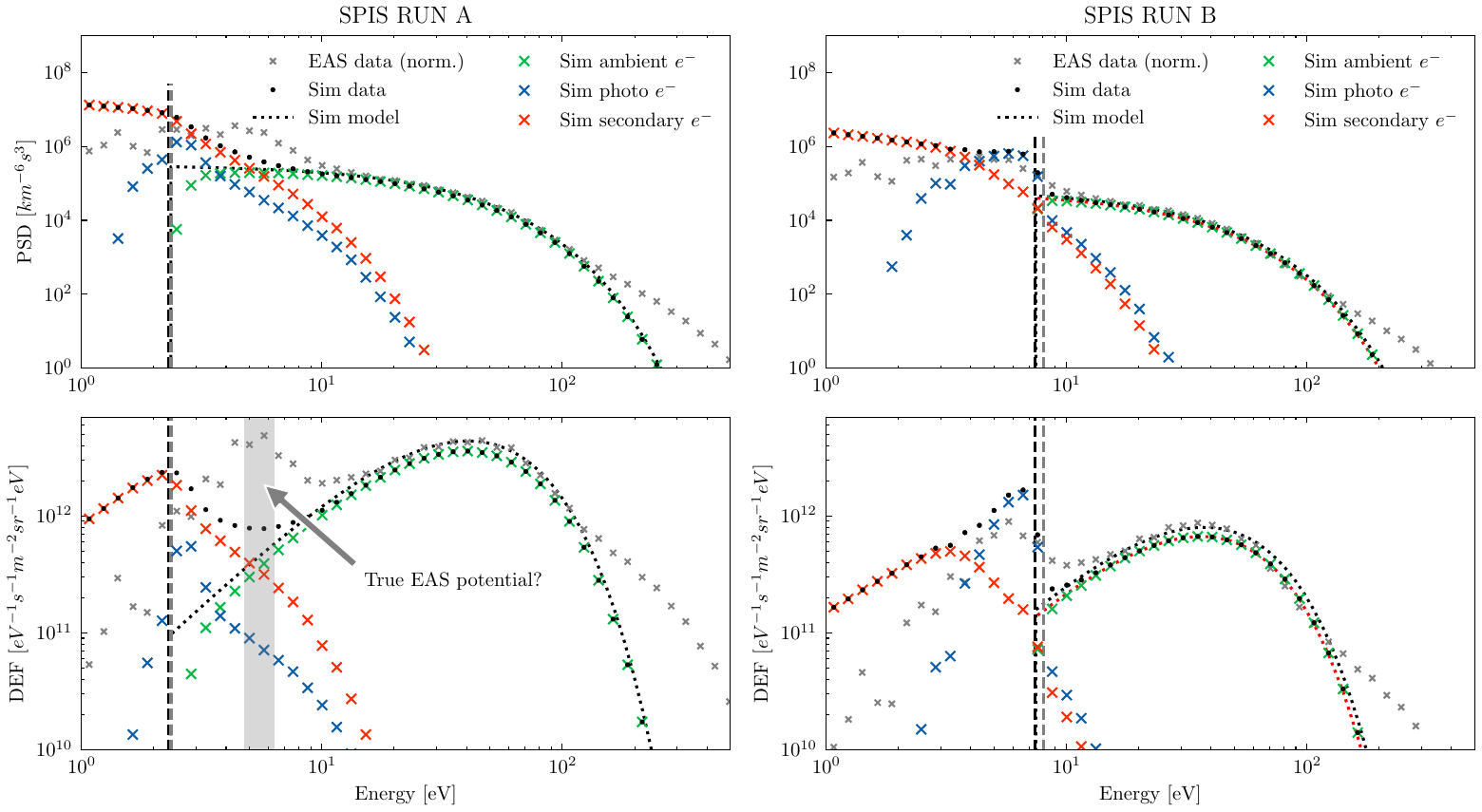}
    \caption{Phase space densities (upper panels) and differential energy flux (lower panels) are shown decomposed into individual contributions of the ambient solar wind (green), secondary (red), and photoelectrons (blue) in comparison to total simulation (black dots) and real SWA-EAS measurements (grey crosses) for both RUN~A (left) and~B (right). Simulated data are over-plotted by the model (dotted line) for ambient background Maxwellian plasma. Simulated spacecraft potential (black dashed line) and the spacecraft potential as measured by RPW is shown for comparison (grey dashed vertical line). Grey-shaded area shows the possible location of the real effective potential of the SWA-EAS detector.}
    \label{fig:vdfA}
\end{figure*}

We analysed the distortion of the simulated distributions measured by the virtual SWA-EAS in more detail using the ability of the SPIS detector to separate individual electron populations, which is not possible for the real SWA-EAS data samples. In Figure~\ref{fig:vdfA}, the total simulated electron energy spectrum (black) is now decomposed into individual contributions from ambient (green), photo (blue), and secondary (red) electron fluxes. The contribution from the ambient electrons qualitatively corresponds, but is lower by about 15\% to the theoretical model response (\ref{eq:fdetector}). An increased deficit with respect to the model is observed at energies directly above the spacecraft potential energy. The distortion of the total distribution from the Maxwellian model for the ambient plasma at low energies is clearly due to the contribution of spacecraft emitted electrons, which populate not only the energies below but also above the energy threshold represented by the spacecraft potential. For simulation run~A, the distortion is dominated by the secondary electron fluxes, while the photoelectron fluxes become comparable (and dominant around the spacecraft potential) for the case of run~B. For both simulation runs, the peak values of the differential energy flux of spacecraft emitted electrons are found around the spacecraft potential energy, which is, however, not the case for the real SWA-EAS samples. The location of the peak values in real measurements is only about 1\,eV below $e\Phi_{SC}$ for sample~B but not for sample~A the real SWA-EAS measurements show the peak values of the differential energy flux around 5-6\,V (see the grey shaded area in the lower left panel of Fig.~\ref{fig:vdfA}) instead of 2-3\,V as observed in the simulated distribution from run~A.

With the SPIS simulations we are not only capable of distinguishing individual electron populations but, for the spacecraft emitted electrons, we are able to even separate different electron fluxes on the SWA-EAS detector based on their source locations across the surface of all spacecraft platform structures. This analysis is presented in Fig.~\ref{fig:simSE} for secondary (left) and photoelectron populations (right) for both simulation runs. Here we divide the main spacecraft structures, namely the main spacecraft body (green), heat shield (purple), solar panels (orange), high-gain antenna (pink), boom (brown), protective baffle (red), and the SWA-EAS sensor itself (blue) to show their contributions to the total fluxes of detected spacecraft emitted electrons. The total distribution of both secondary and photoelectrons is not represented by a single (e.g., Maxwellian) distribution function, but its profile results from a number of different partial distribution functions with various properties changing according to their source locations. 

At lowest energies the measured spacecraft emitted electrons are dominated by secondary emissions from the SWA-EAS detector itself, see the blue curve in the left panels of Fig.~\ref{fig:simSE}. With increasing energy, the highest fluxes of secondary electrons are received from the SWA-EAS baffle, the payload boom, and the main spacecraft body. While secondary electrons emitted from distant spacecraft surfaces are observed up to rather higher energies, secondary electrons emitted from the SWA-EAS surface are reflected and thus measured either by the spacecraft potential (dashed line) or alternatively by the potential barrier (dotted line) created by the negative potential well in the spacecraft wake (see Fig.~\ref{fig:potentials}). The right panels in Fig.~\ref{fig:simSE} show the same analysis for photoelectron emissions. In this case the source locations for the spacecraft electron emissions are clearly restricted to sun-lit surfaces only. The highest fluxes of photoelectrons on the SWA-EAS detector are received from solar panels, with a significant contribution from the heat shield and from the high-gain antenna. The weak contribution from the main spacecraft body is received as a result of small openings in the heat shield, also implemented in our geometrical model of the Solar Orbiter spacecraft. Unlike the secondary electrons, photoelectrons at the SWA-EAS detector are concentrated mostly at energies close to the spacecraft potential. 

\section{Discussion}
\label{sec:dsc} 

\begin{figure*}
    \centering
    \includegraphics[width=1\linewidth]{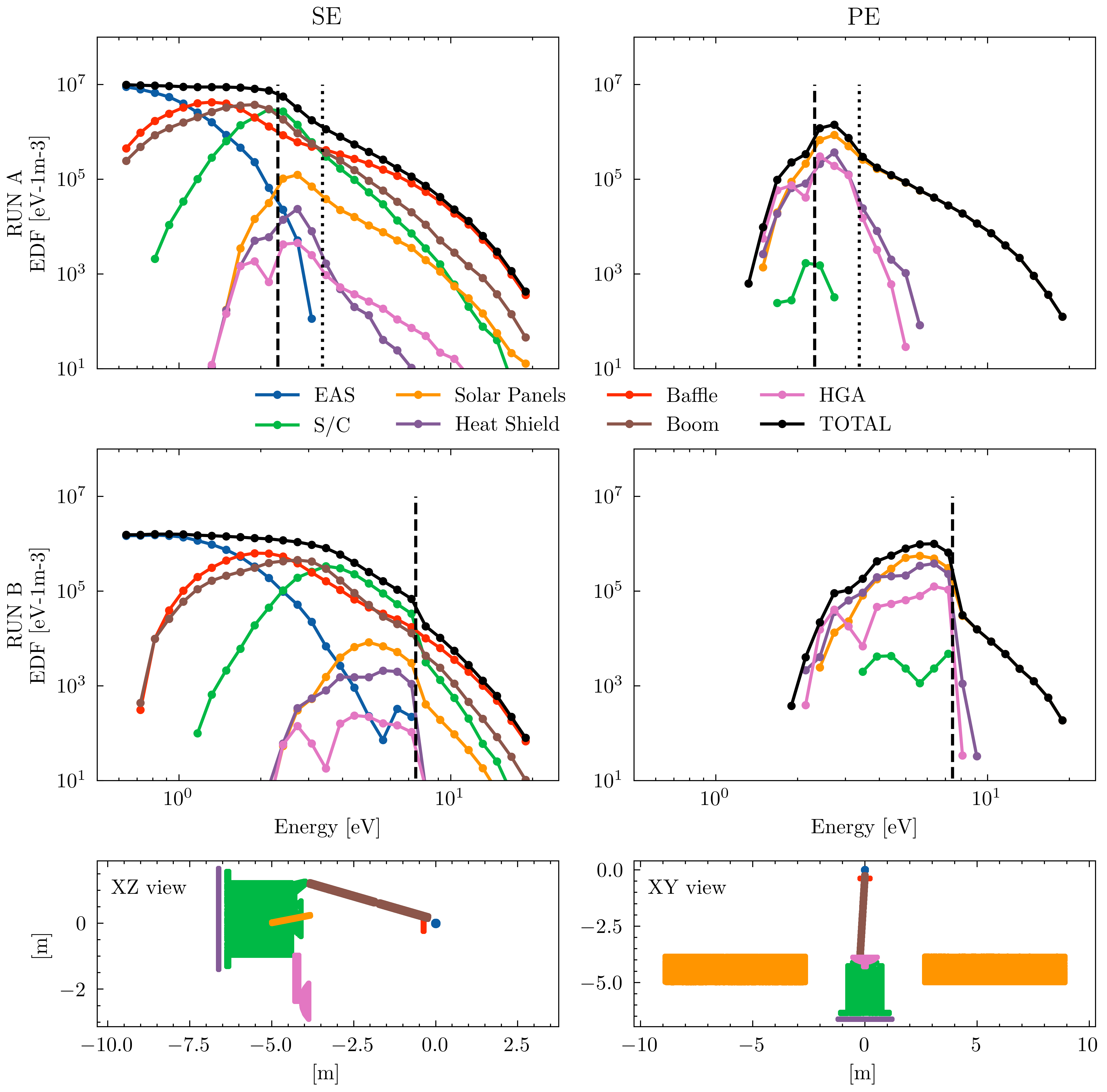}
    \caption{Individual contributions to electron energy distribution function (EDF) as measured by virtual SWA-EAS in the simulation run~A (top row) and run~B (middle row) are shown as a function of the energy for different source surface locations of the secondary emissions (left panels) and photo-emission (right panels): SWA-EAS (blue), spacecraft body (green), solar panels (orange), heat shield (purple), boom (brown), SWA-EAS baffle (red), and HG antenna (pink). The simulated spacecraft potential is plotted with the vertical black dashed line for reference. In the top row (run~A) the dotted line represents the additional potential barrier created in the spacecraft wake. The geometry of the source locations projected to the XZ and YX plane is plotted in the bottom panels.}
    \label{fig:simSE}
\end{figure*}

In the current study we used two SWA-EAS samples we selected to investigate two specific cases, where ({\em i}) the spacecraft potential measured by RPW is relatively small and well below the break in the SWA-EAS electron energy spectrum observed between ambient and spacecraft-emitted electron fluxes, and ({\em ii}) a case of relatively high spacecraft potential where the break location and the spacecraft potential energy are better correlated, cf. sample~A and sample~B in Fig.~\ref{fig:easdata}, respectively. 
The distribution of plasma density in the vicinity of the spacecraft, resulting from the two corresponding simulations run~A and run~B, is similar for both cases (see Fig.~\ref{fig:densities}). In accordance to the main charging processes acting on the spacecraft surface, the charge neutrality is broken near the spacecraft by enhanced concentrations of spacecraft-emitted photo and secondary electrons (see panels in rows b, c, and d), and a distinct ion wake is formed behind the spacecraft structure as a result of the relative drift speed of the spacecraft in the solar wind plasma rest frame (panels in row e). The concentration of secondary electrons is found to be proportional to the ambient plasma density (with the ambient electron temperature being similar in both simulations) so when their densities are plotted normalised to $n_0$ the results of run~A (panels 1-2c) are almost identical to run~B (panels 3-4c). On the other hand, photo-emission rates are by nature independent of the local plasma properties, so that the concentrations of photoelectrons are found stronger relative to $n_0$ in case of run~B with lower initial plasma density (cf. panels 1-2b and 3-4b). Similar properties of these two spacecraft emitted electron populations are also confirmed by virtual SWA-EAS measurements as shown in Figs.~\ref{fig:vdfA} and~\ref{fig:simSE}. Although the photoemission rates typically significantly dominate the secondary emissions, at lower energies, the location of the SWA-EAS detector is mostly dominated by secondary electrons as the detector is mounted in the shadow behind the spacecraft structure, rather far from any sun-lit surfaces emitting high fluxes of photoelectrons.

The geometrical and electrical model of the spacecraft is simplified and not very detailed with respect to the real configuration of the Solar Orbiter. We have verified that in the limiting case of the effective resistivity between nodes R = 0, which corresponds to perfectly conductive spacecraft structures, the simulated spacecraft potential in our model deviates significantly from the RPW measurements. Introducing a finite effective resistance (R = 1\,M$\Omega$) leads to better agreement with the observed potentials, suggesting that imperfect electrical coupling and surface/contact resistances between spacecraft structures play a role. However, the adopted resistance should be regarded as a phenomenological parameter of the model rather than a direct hardware property of the spacecraft. A detailed analysis of structure-to-structure currents is beyond the scope of this study.

The difference between the measured and simulated potential is less than 10\% in case of run~B, and the potentials are almost in agreement in the case of run~A. The spacecraft potential is positive in both cases. Note that at the boundary of the simulation domain, with dimensions higher than those of the local Debye length, the potential is set to zero to represent the unperturbed plasma conditions. As shown already by \cite{Guernsey1970}, two steady state potential distributions can exist in a similar situation for a positively charged photo-emitting plane: the first with a monotonically decreasing potential from the positive value at the surface to zero, and the second with a decreasing potential from the surface until a negative minimum and then with an increase to zero. \cite{Thiebault2004} show by 3D (PIC) simulations that a non-monotonic potential with negative potential minimum can exist all around a positively charged spacecraft with Debye length of the order of the central body radius or more. The negative potential barrier may still surround the entire spacecraft even in the case of an asymmetric illumination pattern with induced photoemission on only one side of the central body. In case of our model, a monotonic profile is found in case of run~B for the higher spacecraft surface potential, while non-monotonic profiles are found in run~A with rather low value of the spacecraft potential (cf. Fig.~\ref{fig:potentials}), in agreement with \cite{Guillemant2013}. 

The performance of the virtual electron detector in our model with respect to the real SWA-EAS instrument was first presented in Fig.~\ref{fig:vdfA_v0}. The plasma parameters, derived from the simulated SWA-EAS measurements by fitting a simple Maxwellian model, show acceptable agreement with the initial plasma conditions, with a slightly underestimated electron density and an overestimated electron temperature. This may possibly be due to the partial blockage of the ambient electron fluxes caused by the spacecraft structure. Clearly, the measured distribution function of ambient electrons in the simulation is found to be slightly below the model corresponding to the background plasma conditions. The deficit in the virtual SWA-EAS measurements of the ambient electrons with respect to the model background increases towards lower energies. This observation can be attributed to the fact that, for electrons with lower energies, the positively charged spacecraft body will have a bigger effective cross-section in comparison to its geometrical dimensions. This naturally leads to the decrease in the total measured electron density, but can result into an increase in the measured electron temperature as the deficit of measured electron fluxes decreases for higher electron energies. The higher temperature provided by the Maxwellian fit of the virtual SWA-EAS measurements may also be attributed to the properties of the model response of the detector, according to the effect of the solar wind bulk speed as discussed in Appendix~\ref{sec:appdet}. The relative error in the underestimated density from the virtual measurements is above 15\%, whereas in our geometrical model the spacecraft structure blocks about 10\% of the full sky. The electron temperature is overestimated by not more than 2\%. Note that the electron energy in the spacecraft frame due to the solar wind bulk speed of roughly 600\,km/s is about 0.7\,eV.

Both the simulated and (real) measured 1D electron energy distributions show similar patterns in their profiles, but with two main differences. First, we model the ambient electrons as a single Maxwellian population only. The supra-thermal (halo) tails observed in the real SWA-EAS measurements are therefore not reproduced in the simulations by the virtual detector. The second and more important difference is observed in the position of the break in the electron energy spectrum, that is the point where the spacecraft-emitted electron fluxes start to dominate the ambient electrons, with respect to the spacecraft potential energy. In both simulation and real in situ measurements, the break is clearly observed above the spacecraft potential threshold, as reported in case of actual SWA-EAS observations by \cite{stverak2025}. However, the first peak of the differential energy flux, caused by the spacecraft-emitted electrons, is found in both simulation runs at the spacecraft potential energy, while this is not the case for the real SWA-EAS and RPW measurements. This raises an important question of whether the real potential on the SWA-EAS detector is identical to the potential on the spacecraft body as provided by RPW (see the grey-shaded area in the lower left panel of Fig.~\ref{fig:vdfA} for illustration). This fact may possibly address the discrepancy in the electron density between the results from RPW and SWA-EAS instruments (see Tab.~\ref{tab:solo_parameter}).
It should be noted that in the simulation model, the SWA-EAS detector is part of the main spacecraft node (see Tab.~\ref{tab:potentials}) and therefore always has the same potential. This may not be the case for the real Solar Orbiter spacecraft and the SWA-EAS detector. An imperfect grounding between individual spacecraft structures may indeed lead to significant differences in the final surface potentials, as shown also in our simulation results for the different electrical nodes (see Tab.~\ref{tab:potentials}).

The importance of geometry and the spatial structure of the potential field is further illustrated by the simulation results in Fig.~\ref{fig:vdfA}. The photoelectron distribution is narrower in energy and has a peak close to the spacecraft potential, while the distribution of secondary electrons from electron impacts is rather broad and spans from $\sim$0 to 20\,eV. The reason for this is in the different source locations of photo- and secondary electron emissions. The photoelectron fluxes are emitted from the heat shield and solar panels far from the SWA-EAS detector and are also directed away from the sensor. Therefore, only relatively fast particles can escape the source location and impact the detector's aperture afterward. Hence, only a rather narrow band of particle energies with convenient trajectories can reach the detector. The secondary electron population is conversely more spread in energy, due to the fact that their source locations are all over the spacecraft: far from the detector as well as very close as the tip of the payload boom or the near baffle shielding the SWA-EAS detector. This effect is demonstrated in Fig.~\ref{fig:simSE} where the photo- and secondary electron distributions are plotted separately for individual source locations of the emissions. The overall intensity of each source depends on its area and proximity to the SWA-EAS detector. The location of the peak intensity in the energy spectrum of the differential energy flux clearly increases towards the spacecraft potential energy with increasing distance of the source to the SWA-EAS sensor.

In general, the simulation results are consistent with and confirm the scenario proposed in \cite{stverak2025}. For electron fluxes measured by the SWA-EAS detector, the spacecraft potential energy represents an effective threshold only for the secondary electrons emitted from the SWA-EAS surface itself and for the ambient electrons (see Fig.~\ref{fig:simSE} and~\ref{fig:vdfA}, respectively). For any spacecraft electrons emitted from other surfaces, real trajectories may still exist with impacting electron energy at the SWA-EAS detector well above the threshold. These particles are first decelerated when leaving the emitting surface and accelerated again towards the SWA-EAS detector by the positive potentials of the spacecraft surfaces. Additional potential barriers may exist due to reported non-monotonic potential profiles. In the case of simulation run~A, such barrier is observed to be created in the wake behind the spacecraft and can effectively reflect escaping spacecraft-emitted electrons back to the SWA-EAS detector, as also demonstrated for a sample trajectory in Fig.~\ref{fig:pe-trajectory}. In case of electrons emitted from the SWA-EAS detector, this potential barrier consequently modifies the effective threshold energy, see the dotted line in the upper left panel of Fig.~\ref{fig:simSE}.

In the present SPIS model, the interplanetary magnetic field is not included. For the near-spacecraft electrostatic environment considered here, this approximation is well justified for thermal solar-wind electrons with energies above 15\,eV, whose gyroradii in typical conditions ($B \approx 40-50$\,nT) are of the order of several tens of meters, i.e., larger than the characteristic size of the spacecraft and its electrostatic sheath. Also, the electric fields associated with the spacecraft potential and local sheath in the order of 1-10\,V/m dominate the acceleration and trajectory bending, while the convective $v \times B$ electric field ($\approx$ few mV/m) produces only minor deflections. Still, for low-energy photo-electrons and secondary electrons with typical energies of a $\approx 1-3$\,eV the corresponding gyroradii may already become comparable to the spacecraft size, and the local magnetic field may therefore introduce modification of their trajectories which can become not fully negligible. However, we note that these effects are expected to mainly influence the fine structure of the emitted-electron transport, while the overall conclusions regarding their net contribution to the measured spectra and the dominant role of electrostatic fields should remain robust.

\section{Conclusions}
\label{sec:con}

The objective of the presented study was to further extend and confirm previous investigations of \cite{stverak2025}. We provided additional analysis, by means of numerical simulations, of the break in the energy distribution function between the cold spacecraft-emitted electrons and the solar wind thermal electrons in relation to the spacecraft potential, as observed in measurements acquired by the SWA-EAS and RPW instruments on board the Solar Orbiter spacecraft. The initial study of \cite{stverak2025} reported, based on real in situ measurements, that the spacecraft potential is, against theoretical assumptions, not directly correlated with the observed energy break that separates the ambient and spacecraft-emitted electron fluxes and that the measurements of ambient thermal electron distributions are still significantly distorted well above the theoretical threshold induced by the spacecraft potential energy. They explained such a discrepancy in the observed energy break by the geometrical configuration of the Solar Orbiter and the SWA-EAS detector itself; however, only by indirect observational proofs.

We have developed and implemented a representative numerical model of the Solar Orbiter spacecraft and its interaction with the streaming solar wind plasma using the Spacecraft Plasma Interaction Software. In the model, we used a simple spherical electron detector, placed at the position of the real SWA-EAS sensors, to provide virtual measurements of 1D integrated energy distribution functions for each of the individual electron populations. With our model, we performed two simulation studies with initial plasma conditions set according to real SWA-EAS observations by the Solar Orbiter at about 0.3~AU for the case of relatively low and high spacecraft potentials and investigated in detail the properties of the energy break in the virtual SWA-EAS measurements. 

Our simulations, in general, have shown a complex structure of the potential field and plasma density around the spacecraft, similar to those presented in previous simulation studies \citep[cf.][]{guillemant2012,guillemant2014,guillemant2017}. For the selected initial plasma conditions, the spacecraft surface is found to have a positive charge with respect to the background plasma, in agreement with the measurements performed by the RPW experiment. The plasma charge neutrality is broken near the spacecraft because of increased concentrations of the spacecraft-emitted electrons, significantly exceeding the background plasma density. The complex potential field in turn leads to significant deviations in trajectories of both ambient and spacecraft-emitted (namely low energy) electrons in the vicinity of the spacecraft. Furthermore, in case of relatively high ambient plasma densities and thus low spacecraft potential, non-monotonic potential profiles were confirmed to exist between the detector and the ambient plasma environment, creating additional potential barriers for the electron trajectories. All of these conditions consequently strongly affect the electron fluxes, as observed by the SWA-EAS detector.

We found a fair qualitative agreement between the simulated virtual measurements and real data observed in situ by SWA-EAS detector. Simulation results, similar to real SWA-EAS measurements, show that contamination by cold electrons emitted from the spacecraft is still observed above the spacecraft potential energy threshold. Detailed analysis of individual simulated electron energy spectra for different electron populations further shows that this contamination above the threshold is caused by cold electron fluxes emitted from distant spacecraft surfaces, as suggested by \cite{stverak2025}. The total flux of the spacecraft-emitted electrons as observed by the SWA-EAS detector is then shown to not behave as a single population, for example, Maxwellian, but as a combination of multiple contributions from different source locations of the emissions with different profiles, still significant for the total electron content in the energy spectrum up to about 15~eV. The spacecraft potential threshold, possibly altered by additional potential barriers in the wake behind the spacecraft, still creates an effective cut-off for measured fluxes of secondary electrons emitted from the SWA-EAS detector itself, but not for emissions from other remaining surfaces of the whole spacecraft structure. The observed break in the electron energy spectrum between the spacecraft and ambient electron populations is moved in this way above the spacecraft potential energy. Consequently, the break itself cannot serve as a reliable estimate of the spacecraft potential for the case of SWA-EAS measurements on board the Solar Orbiter spacecraft.

The relative position of the break in the spectrum with respect to the spacecraft potential differs slightly between the simulations and the observations. The difference was found to be more significant for the case of a higher background plasma density. This disagreement between the simulations and the observations indicates that the potential of the SWA-EAS detector may possibly differ from the potential of the main spacecraft platform. The possible potential difference between the SWA-EAS detector and the spacecraft, not directly confirmed by the present analysis, thus questions the direct use of RPW measurements of the spacecraft potential for on ground SWA-EAS data processing, which can lead to important consequences in the derivation of the unperturbed ambient electron properties from the measured electron velocity distribution functions. 

Future follow-up studies shall therefore focus on any possibilities of alternative spacecraft, or rather real SWA-EAS potential determination. One of the possible procedures involves a detailed directional analysis of measured 3D velocity distributions and an investigation of the energy break as a function of the azimuth and elevation directions. Specific directions not contaminated by cold spacecraft electrons from the distant spacecraft surfaces, may show an energy break in the spectra that may reveal the real potential of the instrument with respect to the background plasma. The numerical model developed within this study can also serve to better understand the errors in estimating the unperturbed plasma properties from the distorted real SWA-EAS observations. Further improvements in the model are of interest, for example, in implementation of the background magnetic field or in increasing the level of detail of the spacecraft geometry and surface material properties. The model can also be extended to include the electrical antennas for direct comparison of spacecraft potential measurements performed by the RPW experiment. 

\begin{acknowledgements}
  Solar Orbiter is a space mission operated by ESA. Solar Wind Analyser (SWA) and Radio and Plasma Waves (RPW) data are derived from scientific sensors which have been designed and created, and are operated by the UCL/MSSL (UK) and LIRA (France) under broad international collaboration. 
  All real in-situ data acquired by Solar Orbiter and used for this study are publicly available at the Solar Orbiter Achive (SOAR).
  The numerical model of the Solar Orbiter used for simulations in this study was developed thanks to the free Space Plasma Interaction Software (SPIS), developed by ONERA and ARTENUM under the support of ESA. 
  This research is primarily supported by the Czech Science Foundation (GAČR) under project No. 23-07334S. The work carried out at UCL/MSSL is supported by UKSA/UKRI grants ST/W001004/1, UKRI919, and UKRI1204. The authors also acknowledge the close and intense collaboration with the SWA-EAS and RPW instrument teams, particularly in discussions related to technical aspects, calibration, and data products. 
\end{acknowledgements}

\bibliographystyle{aa} 
\bibliography{aa58881-26}

\begin{appendix}

\section{VDF formulary}

Assume a general velocity distribution function $f = f(\mathbf{v})$. For the calculation of any moment of the distribution, it is often convenient to use a transformation from the Cartesian to the spherical coordinate system $(\mathbf{v}) \rightarrow (v,\theta,\phi)$, so that
\begin{equation}
    f(\mathbf{v}) d^3\mathbf{v} = f(v,\theta,\phi) v^2 \sin \theta \; dv d\theta d\phi,
\end{equation}
or by substituting the energy $E = \frac{1}{2}mv^2$, with $mv\,dv = dE$,
\begin{equation}
    f(\mathbf{v}) d^3\mathbf{v} = f(E,\theta,\phi) \sqrt{\frac{2E}{m^3}} \sin \theta \, dE d\theta d\phi,
\end{equation}
Let us define $f_v(v)$ as 
\begin{equation}
    \label{eq:fv}
    f_v(v) = \iint_{\Omega} f(v,\theta,\phi) v^2 \sin \theta \, d\theta d\phi,
\end{equation}
or, similarly, $f_E(E)$
\begin{equation}
    \label{eq:fE}
    f_E(E) = \iint_{\Omega} f(E,\theta,\phi) \sqrt{\frac{2E}{m^3}} \sin \theta \, d\theta d\phi
\end{equation}
Assuming an isotropic distribution in the plasma rest frame, $f(v,\theta,\phi) = f(v)$, we can write
\begin{eqnarray}
      f_v(v) & = & 4 \pi v^2 f(v) \\
      f_E(E) & = & 4 \pi \sqrt{\frac{2E}{m^3}} f(E),
\end{eqnarray}
so that $f_E(E) = f_v(v) / (mv)$. From $f(\mathbf{v})$ we can further define the differential number flux $J_N$ 
\begin{equation}
    \label{eq:JN}
    \begin{aligned}
    dJ_N &= v f(\mathbf{v}) d^3 \mathbf{v} \\
    \, &= v^3 f(v,\theta, \phi) \sin \theta \, dv d\theta d\phi \\    
    \, &= \frac{2E}{m^2} f(E,\theta, \phi) \sin \theta \, dE d\theta d\phi,
    \end{aligned}
\end{equation}
and differential energy flux $J_E$ as 
\begin{equation}
\label{eq:JE}
    dJ_E = E dJ_N = \frac{2E^2}{m^2} f(E,\theta, \phi) \sin \theta \, dE d\theta d\phi,
\end{equation}

Now let us assume electrons in space in the plasma rest frame (PRF) with density $n$ at thermal equilibrium with temperature $T$ so the velocity distribution function will take the Maxwellian form 
\begin{equation}
    \label{eq:maxwellPRF}
    f_{PRF}(\mathbf{v}) = n \left( \frac{m_e}{2 \pi k T} \right)^{3/2} \exp \left( -\frac{m_e}{2 k T} v^2\right)
\end{equation}
The distribution function as seen by a spacecraft in its reference frame (SRF), moving with a velocity $\mathbf{u}$ relative to PRF, will now take, according to the Liouville's theorem,
\begin{equation}
    \label{eq:maxwellSRF}
    f_{SRF}(v,\theta,\phi) = n \left( \frac{m_e}{2 \pi k T} \right)^{3/2} \exp \left( -\frac{m_e}{2 k T} (v^2 + u^2 - 2 v u \,\cos \theta) \right),
\end{equation}
knowing that $v_{PRF}^2 = |\mathbf{v}_{SRF} - \mathbf{u}|^2 = v_{SRF}^2 + u^2 - 2 v_{SRF} u \,\cos \theta$, where we have assumed $\mathbf{u}$ to be parallel to the z axis.

In case the spacecraft is charged to a positive potential $\Phi$, the energy of an electron as seen by a particle detector will increase by $e\Phi$, so that 
\begin{equation}
    \label{eq:pottran}
    v_{SRF,\Phi}^2 = v_{SRF}^2 + \frac{2e\Phi}{m_e}.
\end{equation}
Applying the energy shift due to the spacecraft potential while conserving phase-space density by substituting (\ref{eq:pottran}) into (\ref{eq:maxwellSRF}), we get the distribution as seen by the detector to become
\begin{equation}
    \label{eq:maxwellSRFP}
    \begin{aligned}
f_{SRF,\Phi}(v,\theta,\phi) =\;& n \left( \frac{m_e}{2 \pi k T} \right)^{3/2} 
    \exp \left( \frac{e\Phi}{kT}\right) \cdot \\
 \; &\cdot \exp \left( -\frac{m_e}{2 k T} (v^2 + u^2 - 2 u \,\cos \, \theta \sqrt{v^2 - 2e\Phi/m_e}) \right),
    \end{aligned}   
\end{equation}
and all electrons that arrive at the detector will have a velocity $v \ge \sqrt{2e\Phi/m_e}$, cf. \cite{genot2004}.

Finally, substituting (\ref{eq:maxwellPRF})-(\ref{eq:maxwellSRFP}) into (\ref{eq:fv}) leads to
\begin{eqnarray}
    \label{eq:fPRF}
    f_{v,PRF}(v) &=& 4 \pi v^2 n \left( \frac{m_e}{2 \pi k T} \right)^{3/2} \exp \left( -\frac{m_e}{2 k T} v^2\right) \\
    \label{eq:fSFR}
    f_{v,SRF}(v) &=& f_{v,PRF}(v)  \exp \left( -\frac{m_eu^2}{2 k T}\right) \Theta\left( \frac{uvm_e}{kT}\right) \\
    \label{eq:fSFRP}
    f_{v,SRF,\Phi}(v) &=& f_{v,PRF}(v)  \exp \left( -\frac{m_eu^2}{2 k T}\right)  \exp \left( \frac{e\Phi}{kT}\right) \cdot \\ \nonumber 
    & & \cdot \Theta\left( \frac{um_e\sqrt{v^2 - 2e\Phi/m_e}}{kT}\right),
\end{eqnarray}
where
\begin{equation}
    \Theta(x) = \frac{\sinh(x)}{x}
\end{equation}
results from the integration over pitch angle. The factor of 2 arising in Eqs. (\ref{eq:fSFR}) and (\ref{eq:fSFRP}) from the angular integral, 
$$\int_{-1}^{1} \exp(x \mu)\, d\mu = 2\,\sinh(x)/x$$
with $\mu=\cos\theta$, is absorbed into the $4\pi$ prefactor already included in the definition
of $f_{v,\mathrm{PRF}}(v)$, see Eq.~(\ref{eq:fPRF}). 
In limits $u \to 0$ and $\Phi \to 0$, Eqs.~(\ref{eq:fSFR}) and ~(\ref{eq:fSFRP}) reduce to the isotropic PRF expression ~(\ref{eq:fPRF}), providing a consistency check.

\section{Spherical detector}
\label{sec:appdet}
In general, the number of counts detected by any particle detector within a given time interval $dt$ and in the velocity range $d^3\mathbf{v}$ can be, by definition, written as
\begin{equation}
    dC = A(\mathbf{v})dJ_N dt = v f(\mathbf{v}) A(\mathbf{v}) dt d^3\mathbf{v}
\end{equation}
where $f(\mathbf{v})$ is the velocity distribution function and $A(\mathbf{v})$ is the effective surface area of the detector. The total number of particles collected by a spherical detector after integration over all solid angles in the velocity range $dv$ will then be, according to (\ref{eq:fv}), equal to
\begin{equation}
\label{eq:dC}
    dC = v f_v(v) A(v) dt dv
\end{equation}
For a positively charged spherical probe with radius $r_p$ and potential $\Phi$, the effective cross-section $A(v)$ can be derived from the conservation of kinetic energy ($v_p^2 = v^2 + 2e\Phi/m_e$) and angular momentum ($rv = r_pv_p$), so that (cf. \cite{lavraud2016})
\begin{equation}
\label{eq:Av}
    A(v) = \pi r_p^2 \left( 1 + \frac{2e\Phi}{mv^2} \right) = A_p \left( 1 + \frac{2e\Phi}{mv^2} \right).
\end{equation}
Assume that the distribution is constant within $dv$, that is, $f = f_p = const.$, eq. (\ref{eq:dC}) can then be rewritten by use of (\ref{eq:Av}) as
\begin{equation}
\label{eq:countssphere}
\begin{aligned}
dC &= 4 \pi v^3 f A_p \left( 1 + \frac{2e\Phi}{mv^2} \right) dt dv \\
\, &= 4 \pi v_p^3 f_p A_p dv_p dt \\
\, &= v_p f_{v_p} A_p dv_p dt,
\end{aligned}
\end{equation}
where we used the conservation of phase-space density and the relation between $v$ and $v_p$.
Equation~(\ref{eq:countssphere}) directly gives the backward conversion from measured counts to the distribution function of velocity as 
\begin{equation}
    f_{v_p}(v_p) = 4 \pi v_p^2 f(v_p) = \frac{dC}{v_p A_p dt dv_p}
\end{equation}
or similarly for the distribution of energy
\begin{equation}
    \label{eq:fEdC}
    f_{E_p}(E_p) = \frac{dC}{\sqrt{2E/m_e} A_p dt dE_p}
\end{equation}

For the presentation and analysis of both real in situ SWA-EAS samples and simulated measurements from the virtual SPIS detectors we use reduced 1D spectra $f_{FOV}(E_i)$ averaged over all instrument's azimuths $\phi_j$ and elevations $\theta_k$ of the full detector's FOV (see \cite{stverak2025})
\begin{equation}
	f_{FOV}(E_i) = \frac{ \sum_{j = 1}^{n_{\phi}} \sum_{k = 1}^{n_{\theta}} \Delta \Omega_{j,k} f(E_i,\phi_j, \theta_k) }{ \sum_{j = 1}^{n_{\phi}} \sum_{k = 1}^{n_{\theta}} \Delta \Omega_{j,k}},
	\label{eq:fov}
\end{equation}
where $\Delta \Omega_{j,k}$ is the surface on a unit sphere corresponding to the solid angle of the angular bin for the given azimuth and elevation $(\phi_j, \theta_k)$. By use of eq. (\ref{eq:fE}) and (\ref{eq:fSFRP}) we can thus write the model response of a spherical detector to a drifting Maxwellian as
\begin{equation}
    \label{eq:fdetector}
    \begin{aligned}
f_{FOV}(E) =\;& n \left( \frac{m_e}{2 \pi k T} \right)^{3/2} 
    \exp \left( - \frac{E}{kT} -  \frac{m_e u^2}{2 k T} + \frac{e\Phi}{kT}\right) \cdot \\
 \; &\cdot \Theta\left( \frac{u \sqrt{2m_e(E - e\Phi)}}{k T}\right).
    \end{aligned}   
\end{equation}
From eq. (\ref{eq:fdetector}) the electrostatic potential of the detector shifts the obtained energy spectrum by $e\Phi$, while the negative logarithmic slope of the measured distribution ($d\ln(f_{FOV})/dE$) gradually decreases from the asymptotic value $-1/(kT)$ towards $(-1/(kT))( 1 - u^2 m_e / (3kT))$ as $E \rightarrow e\Phi$. The net kinetic energy of the electrons at the detector is increased by $m_eu^2/2$ corresponding to the bulk kinetic energy in the spacecraft reference frame.

\begin{figure}
    \centering
    \includegraphics[width=0.95\linewidth]{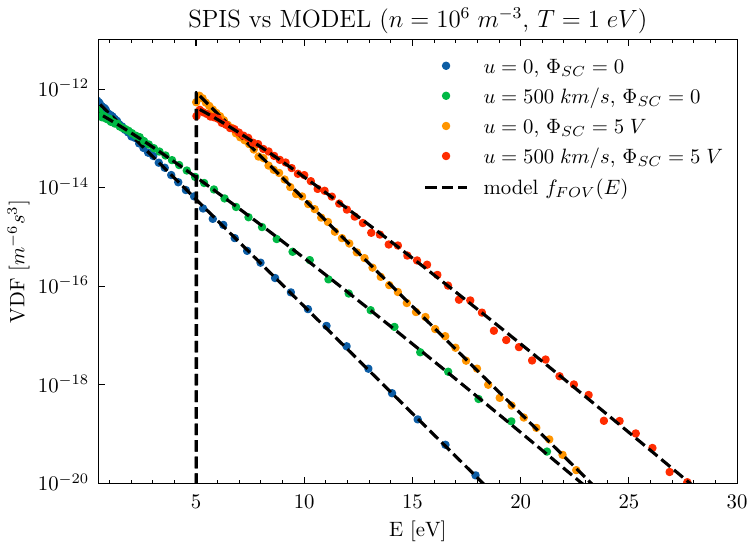}
    \caption{Sample results from a set of simulation runs using SPIS model of a (charged) spherical electron detector immersed in a (drifting) Maxwellian plasma background in comparison to the theoretical response function given by (\ref{eq:fdetector}).}
    \label{fig:fmodel}
\end{figure}

In order to validate the performance of the spherical detector used in our model and to illustrate the effect of the drifting plasma and spacecraft potential on the averaged 1D distribution $f_{FOV}(E)$, we have run a set of SPIS simulations. The simulation setup consists just of the (charged) spherical detector without any S/C body, immersed in a (drifting) Maxwellian plasma. The output from a particle detector in SPIS includes so-called particle list files, which represent statistical samples of particles incident on the detector. Each particle in the list has information on the velocity $v_i$ (or energy $E_i$) of the particle at the surface and the spatial density $n_i$ represented by this particle in the simulation. For a given energy bin $\Delta_E$ the corresponding value of the energy distribution function can be simply derived as
\begin{equation}
    f_{E}(\Delta_E) = \sum_{E_i \in \Delta_E} n_i/ |\Delta_E|.   
	\label{eq:fspis}
\end{equation}
or similarly for the distribution of particle velocities 
\begin{equation}
 f_v(\Delta_v) = \sum_{v_i \in \Delta_v} n_i/ |\Delta_v|.   
\end{equation}
Note that from (\ref{eq:fEdC}) $\sum_{E_i \in \Delta_E} n_i$ is equivalent to $dC / (v A dt)$. The simulated response of the detector for a set of different initial conditions is shown,  compared to the theoretical model given by eq. (\ref{eq:fdetector}), in fig.~\ref{fig:fmodel}. 

\end{appendix}

\end{document}